\documentstyle[11pt,newpasp,twoside,epsfig]{article}
\def\edcomment#1{\iffalse\marginpar{\raggedright\sl#1\/}\else\relax\fi}
\reversemarginpar

\begin{document}
\title{Multi-frequency \& Multi-epoch VLBI study of Cygnus A}
 \author{U. Bach, M. Kadler, T.P. Krichbaum, E. Middelberg, W. Alef, A. Witzel
 and J.A. Zensus} 
\affil{Max-Planck-Institut f\"ur Radioastronomie, Auf dem H\"ugel 69, 53121
Bonn, Germany} 

\begin{abstract}
We present the  first multi-frequency phase-referenced observations of Cygnus A
done with the VLBA at 15 and 22\,GHz. We find a pronounced two-sided jet
structure, with a steep spectrum along the jet and a highly inverted spectrum
towards the counter-jet. The inverted spectrum and the frequency dependent jet
to counter-jet ratio suggest an obscuring torus in front of the counter-jet.
From 14 epochs of 15\,GHz VLBA data we accurately derive the jet and
counter-jet kinematics. For the inner jet ($r \leq 5$\,mas) we measure motions
of $\beta_{\rm app}\approx 0.2-0.5\,h^{-1}$ and on the counter-jet side we
find $\beta_{\rm app}\approx 0.03\pm0.02\,h^{-1}$. We discuss the jet
velocities within the unified jet model. 
\end{abstract}

\section{Introduction}

Cygnus\,A is one of the most powerful radio galaxies at a redshift of
($z=0.057$). It is the archetypical {\sc FR\,II} radio galaxy. In the radio
bands, Cygnus\,A is characterized by two strong lobes separated by 
$\sim$\,2$^\prime $ in the sky. Two highly collimated jets connect the
lobes with the core (Perley et al.\ 1984; Carilli et al.\ 1991). On kiloparsec
scales, the jet is oriented along P.A. $\sim 285^\circ$ and the fainter
counter-jet along P.A. $\sim 107^\circ$. Due to the large inclination of the jet
with respect to the observer, and the correspondingly reduced relativistic
effects, Cygnus\,A is an ideal candidate for detailed studies of its jet
physics, which is thought to be similar to those of luminous quasars (e.g.
Barthel 1989). 

We present and discuss results from multi-frequency VLBI observations at 15, 22
and 43\,GHz and from the first multi-frequency phase referenced observations of
Cygnus\,A done with the VLBA at 15 and 22\,GHz. To complement our data
at 15\,GHz we used ten epochs from the VLBA 2\,cm Survey (Kellermann et al.
1998; Zensus et al. 2002) to analyse the source kinematics with improved
accuracy.

\section{Observations and Data Reduction}

We observed Cygnus\,A in 1996 with the VLBA+Effelsberg at 15, 22, and 43\,GHz,
in 2002 at 5 and 15\,GHz and in 2003 with the VLBA only at 15 and 22\,GHz in
phase-referencing mode. All observations were done in dual circular polarization
and were correlated in Socorro. The data were reduced in the standard manner
using {\sc Aips}. The imaging of the source implying phase and amplitude
self-calibration was done using {\sc Difmap}. 

After imaging we fitted circular Gaussian components to the self-calibrated data
in order to parameterise the source structure. Conservatively, we assume errors
of $\leq10$\,\% in the flux density arising from the uncertainties of the
amplitude calibration and from the formal errors of the model fits. An estimate
for the position error is given by $\Delta r=\frac{\sigma \cdot \Theta}{2 S_{\rm P}}$
(Fomalont\ 1989), where $\sigma$ is the residual noise of the map after the
subtraction of the model, $\Theta$ the width of the component, and $S_{\rm P}$ the
peak flux density. This formula tends to underestimate the error if the peak
flux density is very high or the width of the component is small. In the case of
a small FWHM we used the beam size instead.

\section{Results and Discussion}

To investigate the spectral properties and the kinematics of Cygnus\,A on parsec
scales we cross-identified individual model components along the jet
using their relative separation from each other, their flux density and size. Since the
observations from 2003 were phase-referenced, we compared the
alignment between the two frequencies from the modelfits with those from the
phase-referencing and found similar results. The most likely position 
for the central engine is located between component C3 and J10 (see Fig.~1).
This position shows a slightly inverted spectrum and turned out to be stable in
our kinematical study. It seems to be the same position, which Krichbaum et al.
(1998) already used in their study. 

\begin{figure}[htbp]
\plottwo{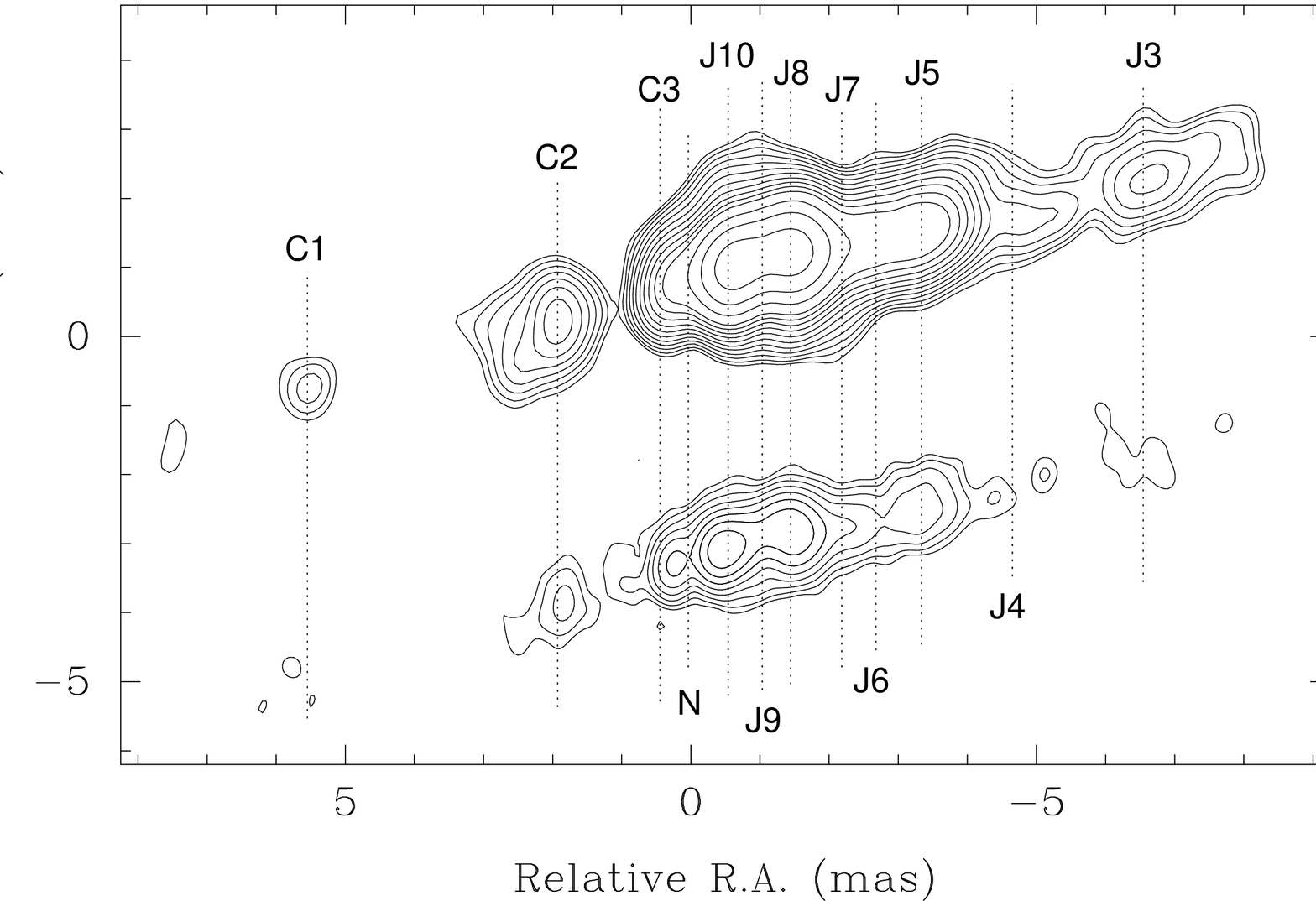}{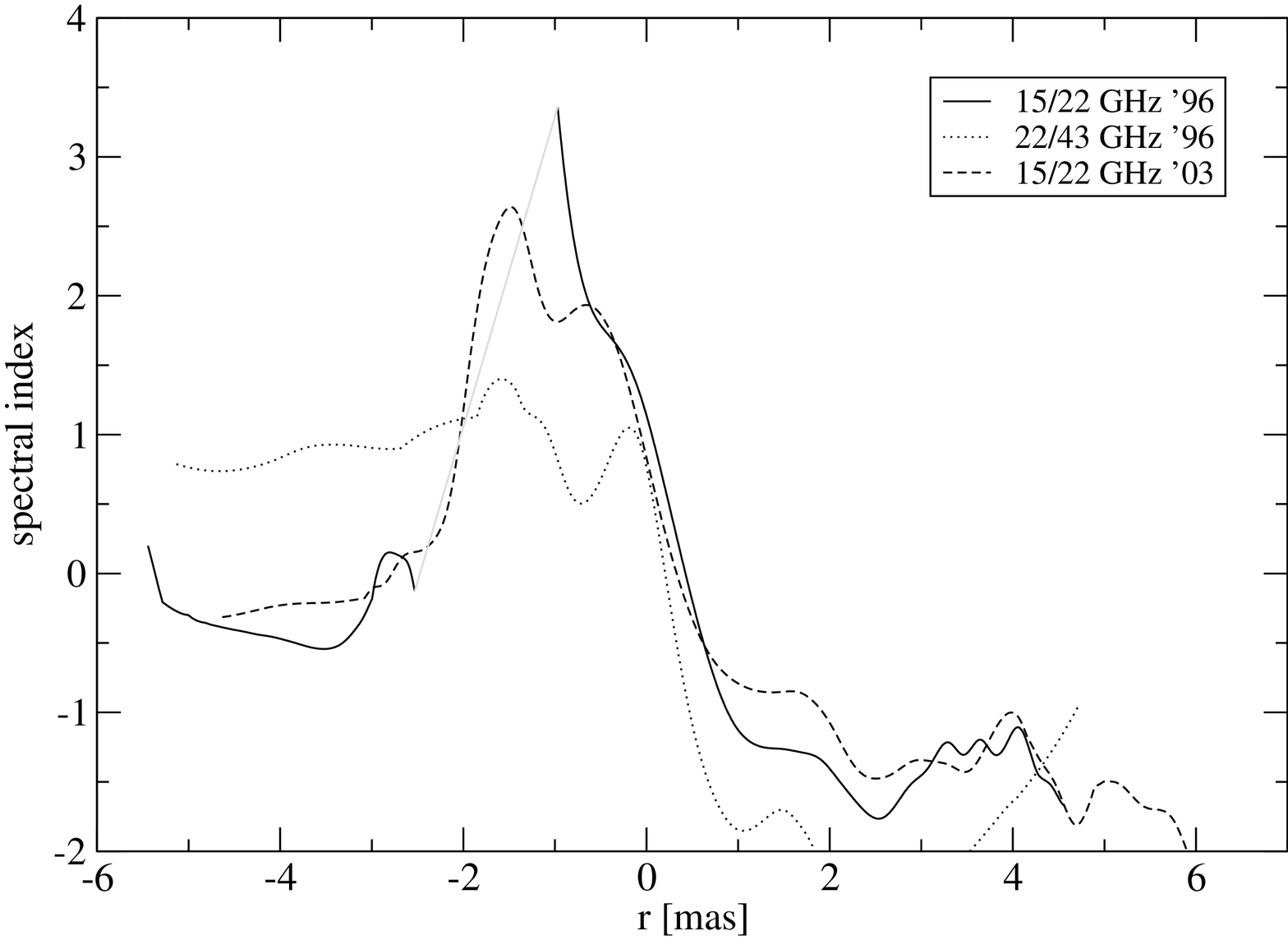}
\caption{\small Left: Phase referenced contour images of Cygnus\,A at 15\,GHz
(top) and 22\,GHz (bottom). The dotted lines indicate the position of the
modelfit components. Right: Profile of the spectral indices along the jet axis,
15/22\,GHz from 1996 and 2003 and 22/43\,GHz from 1996. $r=0$ corresponds to the
position of N 
in the left panel.}
\end{figure}

\subsection{Spectral analysis}

Figure~1 (left) shows the hybrid maps at 15 and 22\,GHz with the
positions of the modelfit components denoted by dotted lines. The images are
registered using phase-referencing to the quasar 2005+403 at a distance of
$1.5^\circ$. On the  right hand side the spectral indices between 15, 22 and
43\,GHz are plotted versus the core distance, assuming N is the core. 

Our analysis reveals that the core spectrum is slightly inverted with a spectral
index of $\alpha_{15/22}\approx \alpha_{22/43}\approx0.5$. The jet shows a steep
spectrum with spectral indices of $\alpha_{15/22}\approx -0.5$ to
$\alpha_{15/22}\approx -1.2$ and $\alpha_{22/43}\approx -1.5$. The counter-jet
has a highly inverted spectrum with $\alpha_{15/22}$ up to 2.5 in the inner part
($r\leq2$\,mas) and a flat spectrum between 15 and 22\,GHz further out. Between 22
and 43\,GHz the counter-jet spectrum is inverted with a spectral index of
$\alpha_{22/43}\approx 1$. The inverted spectrum on the counter-jet side is
likely due to free-free absorption by a foreground absorber. This is supported
by the spectral behavior of the jet to counter-jet flux density ratio (Krichbaum
et al. 1998; Bach et al. 2002). UV spectroscopy (Antonucci et al. 1994) and
optical spectro-polarimetry (Ogle et al. 1997) also show evidence for a hidden
broad line region. According to the unified scheme, this is strong evidence for
an obscuring torus around the central engine and is consistent with the results
from 21\,cm absorption line VLBI (Blanco \& Conway 1996).

\subsection{Kinematics and Geometry}

Figure~2 shows the component positions against time after a careful
identification and linear fits to those, yielding apparent velocities
$\beta_{\rm app}$ as follows. On the jet side the components start 
with 0.1\,mas\,yr$^{-1}$ near the core and accelerate to 0.25\,mas\,yr$^{-1}$ at
larger separations. In Cygnus\,A 1\,mas corresponds to 0.8\,$h^{-1}$\,pc so that
an apparent motion of 0.1\,mas\,yr$^{-1}$ corresponds to $\beta_{\rm
app}=0.26\,h^{-1}$.

\begin{figure}[htbp]
\plottwo{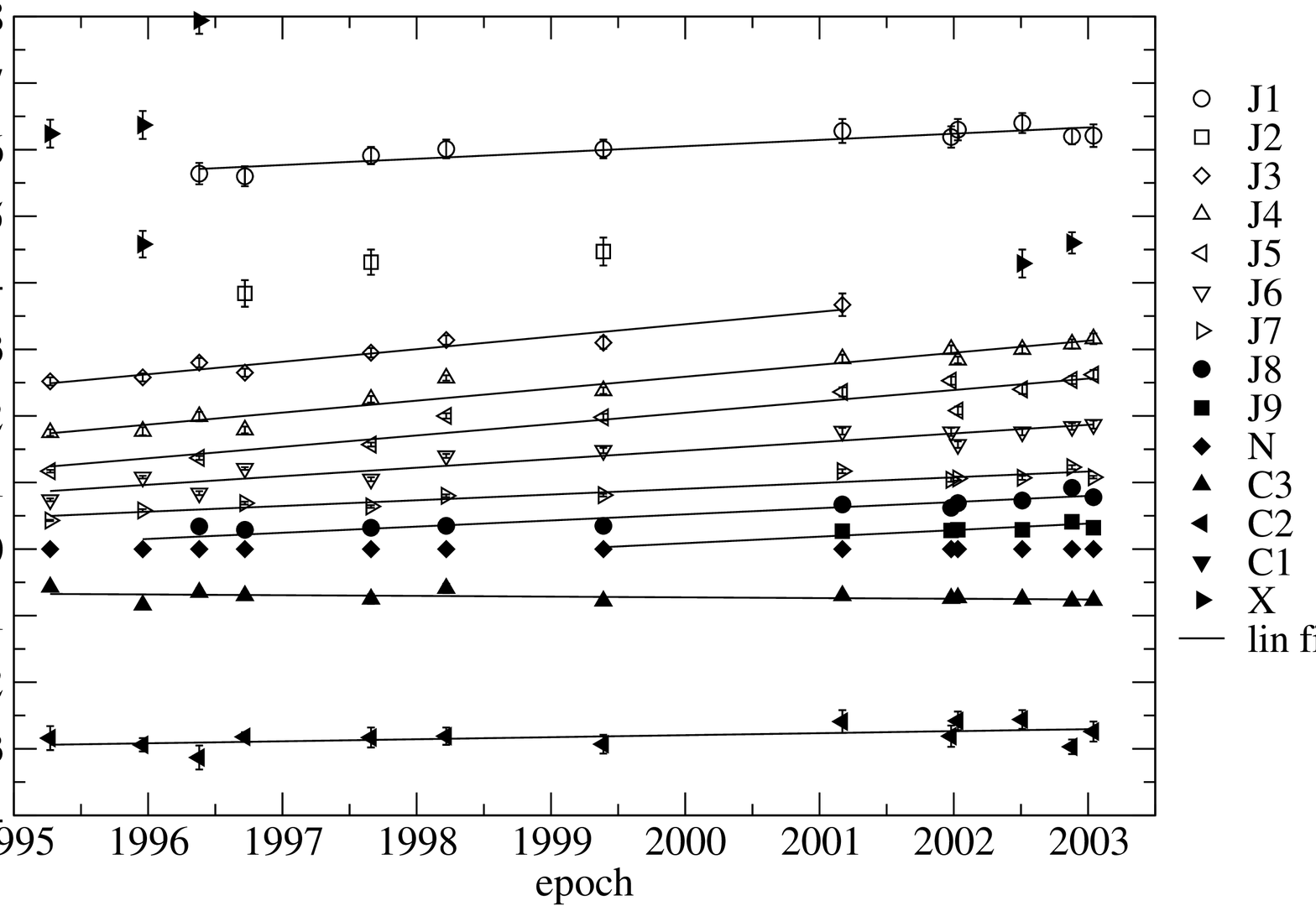}{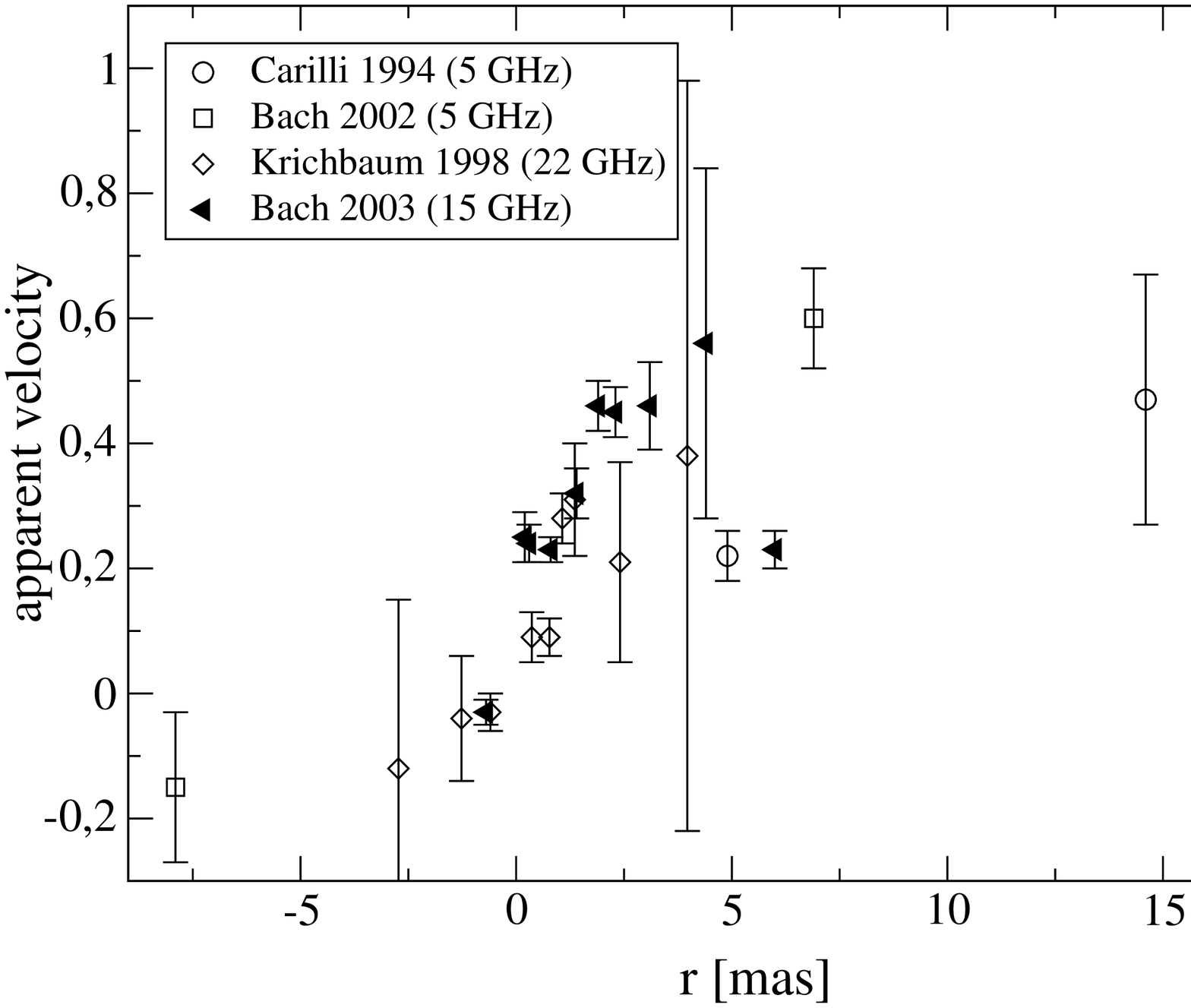}
\caption{\small Left: Component distance at N versus time from 14 epochs
of VLBI data at 15\,GHz. Right: Apparent velocities versus separation
from the core.}
\end{figure}

The situation on the counter-jet side is less clear than on the jet side. The
components are weaker and more extended than on the jet side. The innermost
component C3 shows a marginal motion of $0.010\pm0.007$\,mas\,yr$^{-1}$ and
the next component C2 seems to travel inwards with
$0.02\pm0.01$\,mas\,yr$^{-1}$. In the epochs with the highest quality data,
component C2 consists of two nearly equally bright components. The apparent
inward motion thus might be due to blending effects of these sub-components. A
summary of the component speeds is given in Table~2.
\begin{table}[hbtp]
\centering
\caption{Proper motions of the jet and counter-jet components.}
\begin{tabular}{crr|crr}
\tableline
Comp & \multicolumn{1}{c}{$\mu$ [mas/yr]} &
\multicolumn{1}{c|}{$\beta_{\rm app}$} & Comp & \multicolumn{1}{c}{$\mu$
[mas/yr]} & \multicolumn{1}{c}{$\beta_{\rm app}$}\\  
\tableline
C2 &  $0.02\pm0.01$ &  $0.05\pm0.03$ & J5 &  $0.18\pm0.01$ &  $0.46\pm0.04$\\
C3 & $-0.01\pm0.01$ & $-0.03\pm0.02$ & J4 &  $0.18\pm0.02$ &  $0.45\pm0.04$\\
J9 &  $0.10\pm0.02$ &  $0.25\pm0.04$ & J3 &  $0.18\pm0.03$ &  $0.46\pm0.07$\\
J8 &  $0.09\pm0.01$ &  $0.24\pm0.03$ & J2 &  $0.22\pm0.11$ &  $0.56\pm0.28$\\
J7 &  $0.09\pm0.01$ &  $0.23\pm0.02$ & J1 &  $0.09\pm0.01$ &  $0.23\pm0.03$\\
J6 &  $0.13\pm0.02$ &  $0.32\pm0.04$ & \\
\tableline
\tableline
\end{tabular}
\end{table}
From the right panel in Fig.~2 it seems that the jet components accelerate
as they travel down the jet and reach there maximum speed of $\sim 0.5\,c$ at
a distance of $r\geq2$\,mas. Using $\beta_{\rm app}=\frac{\beta
\sin\theta}{1-\beta \cos\theta}$, which is maximized at $\cot\theta=\beta_{\rm
app}$ and with $\beta_{\rm app}=0.5\,h^{-1}$ one can calculate the lower
limit of the intrinsic velocity 
$\beta_{\rm min}$ of $0.45\,h^{-1}\,c$ and the corresponding angle to 
the line of sight of $\theta=64^\circ$. Indeed, the analysis of the 
jet to counter-jet ratio also favors a large inclination of $80^\circ\pm8^\circ$
(Krichbaum et al. 1998; Bach et al. 2002). 

Due to the small relativistic effects at such conditions, the angle to the line
of sight needs to change by more than $25^\circ$ to explain the observed
velocities by geometrical jet curvature. Since the jet appears to be 
very straight from parsec to kiloparsec scales, it is more likely that we
observe a true acceleration, possibly caused by the collimation of the jet in
the inner 2\,parsec. Alternatively, we might observe a highly stratified jet
with the different velocities belonging to different layers in the jet. This
idea can also explain the absence of detectable motions on the counter-jet side.
Due to the fact that we see the counter-jet from its `back' the emission of the
faster components is beamed away from us and we observe only the slower
velocities of the outer sheath of the counter-jet. 

Bach et al. (2002) detected apparent motion of $\beta_{\rm app}
=0.15\pm0.12\,h^{-1}$ on the counter-jet side at a distance of $r\approx 8$\,mas
from the core at 5\,GHz, but this optically thin component is no more visible at
15\,GHz. From the assumption of a simple K\"onigl jet (K\"onigl\ 1981) and
an inclination of 65 -- 80$^\circ$ we would expect a proper motion of $\beta_{\rm
app} =0.3-0.4\,h^{-1}$ on the counter-jet side. That we do not see these `high'
velocities on the counter-jet points to a more complicated jet structure.

\section{Conclusion}

We carried out the first phase-referencing observations of Cygnus\,A with the
VLBA at 15 and 22\,GHz. The analysis of this data and a multi-frequency
observation from 1996 at 15, 22 and 43\,GHz revealed the spectral properties of
the innermost jet structure in Cygnus\,A. We found a slightly inverted core and
a steep jet spectrum. On the counter-jet side the inner part ($r\leq2$\,mas)
shows a highly inverted spectrum with an 
$\alpha_{15/22}$ of up to 2.5 and $\alpha_{22/43}\approx1$. In the outer regions
the counter-jet shows a flat spectrum between 15 and 22\,GHz and is still
inverted between 22 and 43\,GHz. Together with the frequency dependence
of the jet to counter-jet ratio (Krichbaum et al. 1998; Bach et al. 2002) there
is strong evidence that this is due to free-free absorption by an obscuring
torus. 

The apparent acceleration in the jet and the absence of detectable motions on
the counter-jet side might reflect a more complicated jet structure than that of
the simple K\"onigl jet with a well defined jet flow and
questions the assumption that the jet and counter-jet are intrinsically the
same. The observations could be explained by a stratification of the jet were we
observe different velocities 
sheaths depending on the optical depth and the orientation of the jet. 

\acknowledgments

We thank the group of the VLBA 2cm Survey for providing their data. This work
made use of the VLBA, which is an instrument of the National Radio Astronomy
Observatory, a facility of the National Science Foundation, operated under
cooperative agreement by Associated Universities, Inc. and of the 100\,m
telescope at Effelsberg, which is operated by the Max-Planck-Institut f\"ur
Radioastronomie in Bonn.

\end{document}